\documentclass[pre,11pt,superscriptaddress,preprint]{revtex4-1}

\usepackage{bm,soul}
\usepackage{latexsym}
\usepackage{graphics}
\usepackage{graphicx} 
\usepackage{amsthm} 
\usepackage{amsmath}
\usepackage{wasysym}
\usepackage{amssymb}
\usepackage{enumerate}
\usepackage{slashed}
\usepackage{empheq}
\usepackage[dvips]{epsfig}
\usepackage[colorlinks = true]{hyper ref}
\usepackage{float}  
\usepackage{multirow}
\usepackage{cancel}
\usepackage{bm}
\usepackage[table]{xcolor}
\DeclareMathOperator{\Tr}{Tr}

\newcommand{\mrm}{\mathrm}
\newcommand{\beq}{\begin{equation}}
\newcommand{\eeq}{\end{equation}}
\newcommand{\nn}{\nonumber \\}
  
\newcommand{\hS}{\hat{S}}
\def\bea{\begin{eqnarray}}
\def\eea{\end{eqnarray}}
\hypersetup{
    bookmarks=true,         
    unicode=false,          
    pdftoolbar=true,        
    pdfmenubar=true,        
    pdffitwindow=false,     
    pdfstartview={FitH},    
    pdftitle={My title},    
    pdfauthor={Author},     
    pdfsubject={Subject},   
    pdfcreator={Creator},   
    pdfproducer={Producer}, 
    pdfkeywords={keyword1} {key2} {key3}, 
    pdfnewwindow=true,      
    colorlinks=true,       
    linkcolor=magenta, 
    citecolor=blue,        
    filecolor=magenta,      
    urlcolor=cyan           
}

\begin{document}
\title{Critical behavior of an impurity\\ at the boson superfluid-Mott insulator transition}
\author{Seth Whitsitt}
 \affiliation{Department of Physics, Harvard University, Cambridge, Massachusetts, 02138, USA}
  \author{Subir Sachdev}
 \affiliation{Department of Physics, Harvard University, Cambridge, Massachusetts, 02138, USA}
 \affiliation{Perimeter Institute for Theoretical Physics, Waterloo, Ontario N2L 2Y5, Canada}
 
 \date{\today}
 
\begin{abstract}
We present a universal theory for the critical behavior of an impurity at the two-dimensional superfluid-Mott insulator transition. Our analysis is motivated by a numerical study of the Bose-Hubbard
model with an impurity site by Huang {\it et al.\/} (Phys. Rev. B {\bf 94}, 220502 (2016)), who found an impurity phase transition as a function of the trapping potential. The bulk theory is described by the $O(2)$ symmetric Wilson-Fisher conformal field theory, and we model the impurity by a localized spin-1/2 degree of freedom. We also consider a generalized model by considering an $O(N)$ symmetric bulk theory coupled to a spin-$S$ degree of freedom. We study this field theory using the $\epsilon = 3 - d$ expansion, where the impurity-bulk interaction flows to an infrared stable fixed point at the
critical trapping potential. We determine the scaling dimensions of the impurity degree of freedom and the associated critical exponents near the critical point. We also determine the universal contribution of the impurity to the finite temperature compressibility of the system at criticality. Our results are compared with recent numerical simulations.
\end{abstract}

\maketitle

\section{Introduction}

The quantum phase transition between a superfluid and a Mott insulator in two dimensions represents one of the best studied examples of quantum critical matter, both theoretically and experimentally. The critical properties of this transition are described by a strongly interacting relativistic quantum field theory whose properties have been well-studied in the literature \cite{FWGF89,ssbook}. This phase transition can be realized experimentally using cold atoms trapped in optical lattices, providing greater access to its properties \cite{G02Nature,Chin11,Endres12}. 

In this paper, we study the superfluid-insulator transition in the presence of an impurity degree of freedom, motivated by recent numerical work by Huang {\it et al.\/} \cite{HCDS16} of the lattice Bose-Hubbard
model. In their study, they model the presence of an impurity in terms of a trapping potential, representing the attachment of charge to the impurity. They find the emergence of scale-invariant behavior for a critical value of the trapping potential, suggesting the emergence of a new universality class associated with the impurity degree of freedom.

Models of impurities coupled to a bulk theory were considered in References \onlinecite{SS99,AMS00}. Furthermore, a model of impurities coupled to a bulk interacting critical theory was investigated in Refs.~\onlinecite{SBV99,VBS00,SS00}. The latter model describes the effect of impurities coupled to quantum antiferromagnets close to their critical point. In that work, the impurities are represented by a localized spin degree of freedom which coupled to the bulk quantum field theory, and a stable interacting fixed point was found perturbatively in the $\epsilon = 3 - d$ expansion. This novel impurity-driven critical behavior led to new observables associated with the impurity degree of freedom.

Here we take a similar approach in studying the superfluid-insulator transition coupled to impurities. We will argue for the particular form of an impurity-bulk interaction to model the critical behavior, and study the resulting theory in the $\epsilon$ expansion. Working with a slightly generalized model, we will find an interacting fixed point, and calculate the new critical exponents associated with the theory. We will also determine the universal dependence of the finite temperature compressibility on the impurity degree of freedom. The exponents and the compressibility can be related to those calculated numerically in Refs.~\onlinecite{HCDS16,chen17}.

Our paper is arranged as follows. In Section \ref{sec:model} we discuss the microscopic model of Ref.~\onlinecite{HCDS16}, and argue for the form of the universal quantum field theory describing its universal properties. We set up the form of the $\epsilon$ expansion of a generalized form of the theory. Section \ref{sec:renorm} describes how the diagrammatic expansion of the model is constructed, and gives the expansion to two-loop order. We give a summary of the renormalization group equations in Section \ref{sec:rgsumm}, and give our predictions for the critical exponents of the model. In Section \ref{sec:compress}, we determine the universal contribution of the impurity to the finite temperature compressibility of the model, and we conclude in Section \ref{sec:summ}.

\section{The model}
\label{sec:model}

\subsection{Continuum field theory}

We seek the critical theory describing the microscopic model studied numerically in Ref.~\onlinecite{HCDS16}.  This is given by
\beq
\mathcal{H}_1 = \sum_{\langle ij \rangle} b^{\dagger}_i b_j + \frac{U}{2} \sum_{i} n_i \left( n_i - 1 \right) - \mu \sum_i n_i + V n_0
\label{microH}
\eeq
where $b^{\dagger}_i$ is a boson creation operator on site $i$, $\langle \cdots \rangle$ denotes nearest-neigbors, and $n_i \equiv b^{\dagger}_i b_i$. The model is studied at constant density with unit filling fraction, where a bulk critical point between a superfluid and insulating state is known to exist at the values $U_c = 16.7424(1)$ and $\mu_c = 6.21(2)$ \cite{CSPS08,SKPS11}. For $V=0$, it is known that the bulk transition is described by the relativistically-invariant $O(2)$-symmetric Wilson-Fisher conformal field theory \cite{FWGF89}, given by the Hamiltonian
\beq
\mathcal{H}'_{\phi} = \int d^3x \left\{ \frac{ \pi_{\alpha'}^2 + c^2 \left(\nabla \phi_{\alpha'}\right)^2 + s_c \ \phi_{\alpha'}^2}{2} + \frac{u_0}{4!} \left(\phi_{\alpha'}^2\right)^2 \right\}
\label{bulkH}
\eeq
where the index runs from $\alpha' = 1,2$. The coupling $s_c$ has been fine-tuned to its critical value, and $u_0$ flows to a universal value in the infrared. The fields $\phi_{\alpha'}(x,t)$ and $\pi_{\alpha'}(x,t)$ represent the bulk order parameter and its canonical conjugate respectively, obeying the commutation relation
\beq
[\phi_{\alpha'}(x,t),\pi_{\beta'}(x',t)] = \delta_{\alpha' \beta'}\delta^d(x - x')
\eeq 
The velocity scale $c$ depends on microscopic details of the system, and will henceforth be set to unity.

In Ref.~\onlinecite{HCDS16}, it was found that the addition of the impurity potential $V$ leads to new critical behavior. As the potential is turned on, it is found that charge is either depleted or concentrated at the origin depending on the sign of $V$, with a density profile characterized by a half-integer charged core and a half-integer charged halo located at a radius $\xi_{h}$ from the origin.

At a critical value of $V$, the halo diverges to infinity, indicating the onset of scale invariance. If the coupling $V$ continues to increase, the charge of the halo changes sign and contracts back to the origin, so this is a transition between a system with total charge $Q$ and $Q \pm 1$. The radius of the halo is observed to have the universal behavior
\beq
\xi_h \propto |V - V_c|^{-\nu_z}
\eeq
with $\nu_z = 2.33(5)$ \cite{HCDS16}.

In seeking the critical theory, we need to couple the bulk Hamiltonian Eq.~(\ref{bulkH}) to a field describing the impurity degree of freedom. This theory retains the $O(2)$ invariance. We claim that the correct impurity coupling is given by
\beq
\mathcal{H}_{\mathrm{imp}} = -\gamma_0 \left[ \phi_{1}(x = 0) \hS_{x} + \phi_{2}(x = 0) \hS_{y} \right] + h_z \hS_z
\label{spincoup}
\eeq
where $\hS_{\alpha}$ represents a spin-1/2 degree of freedom defined at $x=0$; a spin $S=1/2$ impurity model has also been proposed and studied independently by Chen {\it et al.\/} \cite{chen17}. We also note that a scalar-spin interaction of this form was studied in a different context by Zar\'{a}nd and Demler \cite{ZD02}.
Here, the two couplings $\gamma_0$ and $h_z$ are both relevant in $d=2$. The $O(2)$ symmetry of the impurity is generated by $\hS_z$, and at $h_z = 0$, there is an exact two-fold degeneracy between the $\hS_z = \pm 1/2$ states, which reproduces the two-fold degeneracy of the microscopic theory at the critical impurity potential $V = V_c$ between the different charge sectors. We will argue below that the coupling $\gamma_0$ flows to a universal value which controls the critical behavior of the impurity degree of freedom.

Our analysis will also consider the case where the impurity $\hS_\alpha$ has a generic spin $S$. This corresponds to possible multicritical points where $2S+1$ states become degenerate at the impurity. In the Bose-Hubbard model, we would have to tune $2S$ couplings to achieve this. In the field theory, the $2S$ relevant couplings correspond to the operators $\hat{S}_z^p$, with $1 \leq p \leq 2 S$. We will only consider the scaling dimension of the $p=1$ operator here.

\subsection{Expansion in $\epsilon$}

We will work with a generalization of the above theory, given by
\beq
\mathcal{H} = \mathcal{H}_{\phi} -\gamma_0 \phi_{\alpha'}(x=0) \hat{S}_{\alpha'}
\label{eq:model}
\eeq
Here, the first term is the Hamiltonian for the $O(N)$-symmetric scalar field theory in $d$ spatial dimensions,
\beq
\mathcal{H}_{\phi} = \int d^dx \left\{ \frac{ \pi_{\alpha}^2 + \left(\nabla \phi_{\alpha}\right)^2 + s_c \ \phi_{\alpha}^2}{2} + \frac{u_0}{4!} \left(\phi_{\alpha}^2\right)^2 \right\}
\eeq
We use the notation where unprimed indices run from $\alpha = 1,2, ..., N$, while primed indices only take the values $\alpha' = 1,2$. Summation is implied over repeated indices, and it is understood that $\phi_{\alpha}^2 = \phi_{\alpha} \phi_{\alpha}$. The operators $\hS_{\alpha}(t)$ satisfy the $SU(2)$ algebra,
\bea
[\hS_{\alpha},\hS_{\beta}] &=& i \epsilon_{\alpha \beta \gamma} \hS_{\gamma} \nn[.3cm]
\Tr\left( \hS_{\alpha} \hS_{\beta} \right) &=& \frac{1}{3}(2S+1)S(S+1) \delta_{\alpha \beta}
\label{eqn:spincom}
\eea
where the spin operator only takes the values $\alpha = 1,2,3$. We continue to label the $1-2$ directions with primed indices, and refer to the third direction as the $z-$direction. We note that the total Hamiltonian in Eq.~(\ref{eq:model}) has $O(2)\times O(N - 2)$ symmetry. Here we will allow arbitrary values of spin, $S$, and give results for $S = 1/2$ at the end of the calculation. Although the operator $\hS_{z}$ does not appear in this Hamiltonian, it has nontrivial correlations in the interacting theory due to the commutation relations. Its scaling dimensions will then determine the critical exponent associated with perturbing this theory by a term $h_z \hS_{z}$.


We will study this system in the $\epsilon = 3 - d$ expansion. We will use the minimal subtraction renormalization scheme of Ref.~\onlinecite{zinn2002}, where $s_c = 0$ and the bare fields and interaction strength are replaced by
\bea
\phi_{\alpha} &=& \sqrt{Z} \phi_{R \alpha} \nn
u_0 &=& \frac{\mu^{\epsilon} Z_4}{S_{d+1} Z^2} g
\label{eqn:rdef}
\eea
Here, $\mu$ is an arbitrary energy scale, $g$ is a dimensionless coupling constant, and 
\beq
S_d = \frac{2}{\Gamma(d/2) (4 \pi)^{d/2}}
\eeq
is a convenient phase factor. To leading order in $g$, the renormalization constants are given by
\bea
Z &=& 1 - \frac{(N+2)}{144 \epsilon}g^2 \nn[.3cm]
Z_4 &=& 1 + \frac{(N+8)}{6 \epsilon}g + \left( \frac{(N+8)^2}{36 \epsilon^2} - \frac{(5N + 22)}{36 \epsilon} \right) g^2
\label{eqn:rcond}
\eea
The beta function follows immediately from Eqns.~(\ref{eqn:rdef}) and (\ref{eqn:rcond})
\beq
\beta_{g} \equiv \mu \frac{dg}{d \mu}\bigg|_{u_0} = - \epsilon g + \frac{(N+8)}{6} g^2 - \frac{(3N+14)}{12}g^3
\eeq
from which we determine the bulk fixed point by finding the value of $g$ where the beta function vanishes:
\beq
g^\ast = \frac{6 \epsilon}{(N+8)}\left[ 1 + \frac{3(3N+14)}{(N+8)^2}\epsilon \right]
\eeq

The addition of a localized bulk-impurity interaction cannot significantly alter the bulk correlation functions, so the above results also hold for the full theory $\mathcal{H}$. However, we must now consider the renormalization of the impurity operators and their interaction with the bulk order parameter. We define the constants
\bea
\hS_{\alpha'} &=& \sqrt{Z'} \hS_{R \alpha'} \nn
\hS_{z} &=& \sqrt{Z_z} \ \hS_{R z} \nn
\gamma_0 &=& \frac{\mu^{\epsilon/2} Z_{\gamma}}{ \sqrt{Z Z' \tilde{S}_{d+1}} } \gamma
\eea
Here, $\gamma$ is another dimensionless renormalized interaction, and we have introduced another convenient phase factor
\beq
\tilde{S}_{d} = \frac{\Gamma(d/2 - 1)}{4 \pi^{d/2}}
\eeq
In terms of the above constants, we find that the impurity beta function is given by
\beq
\beta_{\gamma} \equiv \mu \frac{d \gamma}{d \mu}\bigg|_{u_0,\gamma_0} = - \frac{\frac{\epsilon}{2} \gamma + \gamma \beta_g \frac{d}{dg} \log \left( Z_{\gamma}/\sqrt{Z Z'} \right)}{1 + \gamma \frac{d}{d\gamma} \log \left( Z_{\gamma}/\sqrt{Z Z'} \right)}
\label{eqn:gammabeta}
\eeq

One major result of this paper is the determination of the beta function to two-loop order, from which we find an infrared fixed point at a critical value of $\gamma^{\ast}$ which is perturbative in $\epsilon$. The major observables associated with this fixed point are the universal decay of the spin operators. We introduce the anomalous dimensions,
\bea
\left\langle \hat{S}_{\alpha'}(t) \hat{S}_{\alpha'}(0) \right\rangle \sim \frac{1}{t^{\eta'}} \nn[.3cm]
\left\langle \hat{S}_{z}(t) \hat{S}_{z}(0) \right\rangle \sim \frac{1}{t^{\eta_z}}
\eea
where algebraic decay is forced by scale invariance, and the exponents are given by
\bea
\eta' &=& \beta_{\gamma} \frac{d}{d \gamma} \log Z' + \beta_{g} \frac{d}{d g} \log Z' \nn
\eta_z &=& \beta_{\gamma} \frac{d}{d \gamma} \log Z_z + \beta_{g} \frac{d}{d g} \log Z_z
\label{eqn:andim}
\eea
These anomalous dimensions, which are twice the scaling dimension of the spin operators, are new data associated with the universality class of this phase transition. 

Once the anomalous dimension of $\hS_z$ is determined, we can also determine the critical exponents associated with perturbing the critical theory. The leading relevant perturbations to Eq.~(\ref{eq:model}) are given by
\bea
\Delta\mathcal{H}' = h' \hS_{\alpha'} \nn
\Delta\mathcal{H}_z = h_z \hS_{z}
\eea
for any of the three $\hS_{\alpha}$. This perturbation will introduce a large timescale $\xi$ characterizing an exponential decay of the spin correlation functions, and by scaling arguments, it is straight-forward to show that
\bea
\xi' = |h'|^{-\nu'} \nn
\xi_z = |h'|^{-\nu_z}
\eea
where
\bea
\nu' = \frac{1}{1 - \eta'/2} \nn
\nu_z = \frac{1}{1 - \eta_z/2}
\label{nudef}
\eea
Here, exponent $\nu_z$ corresponds to the critical exponent defined in the microscopic model above. 

\section{Renormalization}
\label{sec:renorm}

We determine the renormalization parameters above using bare perturbation theory. In particular, we will work in imaginary time $\tau$, and compute the following correlation functions to two-loop order:
\bea
\mathcal{G}'(\tau) \delta_{\alpha' \beta'} &=& \left\langle \hS_{\alpha'}(\tau) \hS_{\beta'}(0) \right\rangle \nn[.3cm]
\mathcal{G}^z(\tau) &=& \left\langle \hS_{z}(\tau) \hS_{z}(0) \right\rangle \nn[.3cm]
\mathcal{V}(x,\tau) \delta_{\alpha' \beta'} &=& \left\langle \phi_{\alpha'}(x,\tau) \hS_{\beta'}(0) \right\rangle
\label{eqn:correlators}
\eea
All correlation functions are understood to be imaginary time-ordered, we take $\tau>0$, and a trace is taken over the spin indices. This calculation will result in divergences in the form of poles in $\epsilon$, but we choose the constants $Z'$, $Z_3$, and $Z_4$ such that these poles cancel when the correlation functions are expressed in terms of renormalized operators and couplings.

Due to the nontrivial commutator in Eq.~(\ref{eqn:spincom}), the perturbative expansion for these correlation functions does not obey Wick's theorem, nor do disconnected diagrams cancel. We must expand the numerator and denominator of the correlation functions separately as a series in $u_0$ and $\gamma_0$, and by carefully keeping track of the time-ordering of the spin operators we can obtain the desired correlation functions. This procedure can be represented by a form of diagrammatic perturbation theory developed in Ref.~\onlinecite{VBS00}. 

We first write the correlation function of the interacting theory in terms of free correlators, where the free part of our theory is the quadratic part of $\mathcal{H}_{\phi}$.
\bea
\langle \mathcal{O} \rangle = \frac{\left\langle \mathcal{O}  e^{-\beta \mathcal{H}_I } \right\rangle_0}{\left\langle e^{-\beta \mathcal{H}_I } \right\rangle_0}
\label{eq:expv}
\eea
We introduce a finite inverse temperature $\beta$ as an intermediate step. The Hamiltonian which appears on the right-hand side is the interaction Hamiltonian,
\beq
\mathcal{H}_I = \frac{u_0}{4!} \int d^dx \left(\phi_{I \alpha}(x)^2\right)^2 - \gamma_0 \phi_{I \alpha'}(x = 0) \hS_{\alpha'}
\eeq
The operators $\phi_{I \alpha}$ are the familiar interaction representation of our original bosonic fields (the interaction and Schr\"odinger representations of  $\hS_{\alpha}$ are equivalent in our model). Then we expand the exponentials in the numerator and denominator, and the expectation values break into simple products of bosonic correlators and spin correlators. The bosonic operators obey Wick's theorem, so we obtain integrals over products of the free finite-temperature bosonic Green's function:
\beq
D_T(x,\tau) = \langle \phi_{\alpha}(x,\tau) \phi_{\beta}(0,0) \rangle_0
\eeq
However, the time-ordering over spin expectation values will result in a corresponding time-ordering over dummy integration variables. 

We represent the imaginary time-ordered expectation value of an arbitrary operator $\langle A \rangle_0$ with the following diagrammatic rules:
\begin{itemize}
\item Every diagram contains a single directed loop along which imaginary time runs periodically from $0$ to $\beta$, represented by a full line.
\item External factors of $\hS_{\alpha}(\tau)$ contained in $A$ are represented by open circles placed on the directed loop at the appropriate external value of $\tau$.
\item External factors of $\phi_{\alpha}(\tau,x)$ contained in $A$ are represented by open boxes which are placed outside of the directed loop.
\item Factors of the interaction $\gamma_0$ are represented by closed circles placed on the directed loop, and a bosonic propagator always emerges from this vertex. 
\item Factors of the interaction $u_0$ are represented by a filled square, which connects to four bosonic propagators.
\item Internal bosonic propagators connecting vertices placed at $(x_i,\tau_i)$ and $(x_j,\tau_j)$ give a factor of $D_T(x_i - x_j,\tau_i - \tau_j)$, and we integrate over all internal $x_i$ and $\tau_i$. However, the ordering of all the $\tau_i$'s appearing on the directed loop must be kept in determining the integration region.
\item We trace over the spins along the directed line. If there are no spin operators inserted, this is interpreted at $\Tr \mathbb{I} = (2S+1)$.
\end{itemize}
We obtain the correction to $\langle A \rangle$ at a given order of $u_0$ and $\gamma_0$ by writing down all possible diagrams which obey the above rules and have the correct number of interaction vertices, and then sum them. We will demonstrate how to apply these rules in detail for the relatively simple one-loop case, before giving the full two-loop results.

\subsection{Spin-spin correlation function}
\label{sec:2pnt1}
\begin{figure}
\includegraphics[width=15cm]{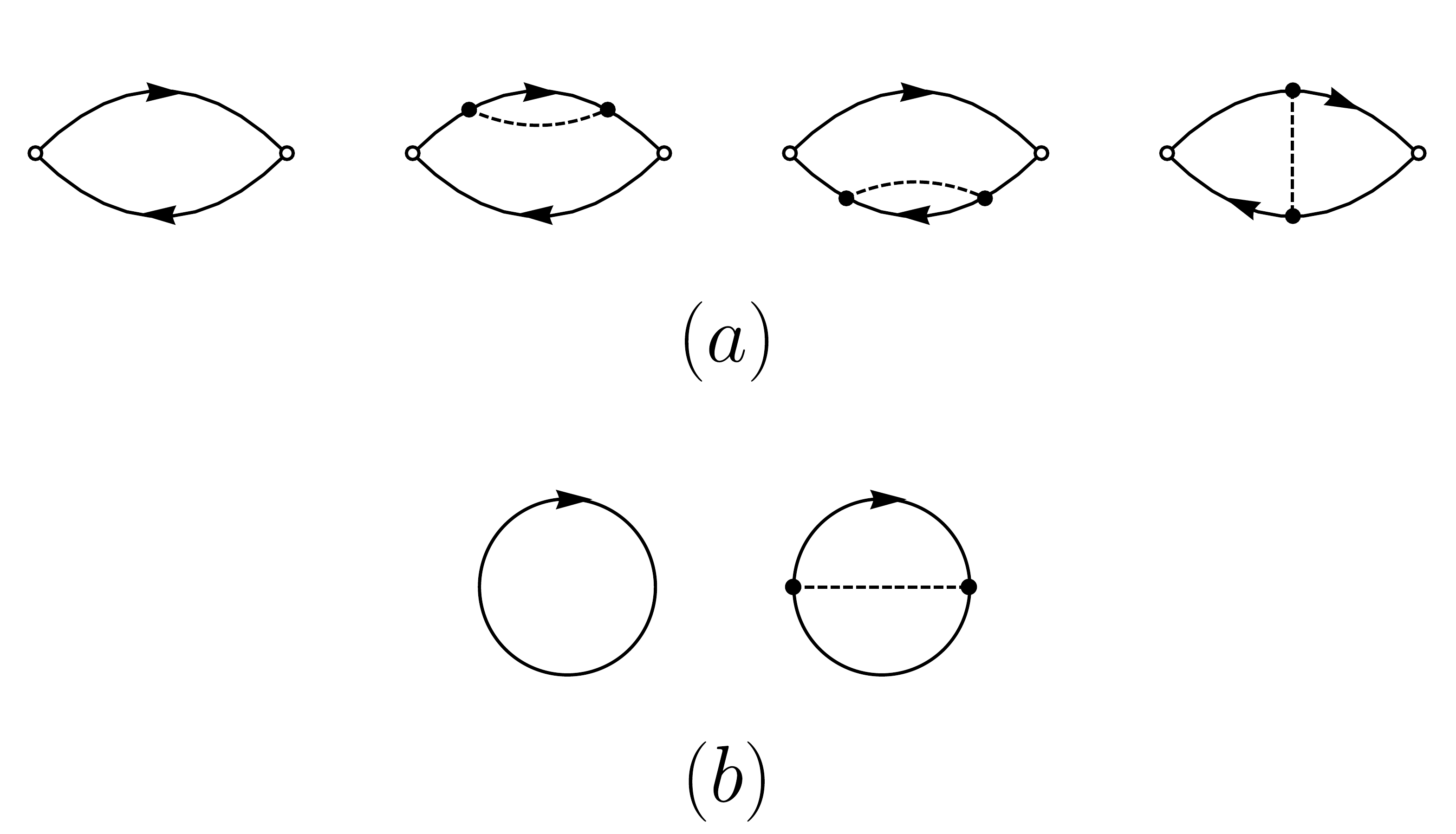}
\caption{The diagrammatic expansion for the spin-spin correlation function at one-loop, using the Feynman rules specified in the Section \ref{sec:renorm}. The diagrams contributing to the numerator and denominator of the correlation function are pictured in (a) and (b) respectively. As described in the main text, the integrals contributing to the numerator and denominator can be combined, so that we only need to keep track of differing spin traces.}
\label{fig:2pnt1}
\end{figure}

We show the lowest-order diagrams contributing to the spin-spin correlation functions in Fig.~\ref{fig:2pnt1}.(b). Below we will evaluate spin traces using the identities enumerated in Appendix \ref{traces}. We first write out the diagrams in the denominator, obtaining from the above rules
\beq
\mathcal{Z} = (2 S + 1) + \Tr\left( \hS_{\alpha'} \hS_{\alpha'} \right) \gamma_0^2 \int_0^{\beta} d\tau_1 \int_{\tau_1}^{\beta} d \tau_2 D_T(\tau_1 - \tau_2) + \cdots
\eeq
We then rewrite this expression for reasons which will become clear shortly:
\bea
\mathcal{Z} &=& (2 S + 1) + \ (2 S + 1)\frac{2 S (S+1)}{3} \gamma_0^2 \bigg[ \int_0^{\tau} d\tau_1 \int_{\tau_1}^{\tau} d \tau_2 + \int_{\tau}^{\beta} d\tau_1 \int_{\tau_1}^{\beta} d \tau_2 \nn
&& + \ \int_0^{\tau} d\tau_1 \int_{\tau}^{\beta} d \tau_2 \bigg] D_T(\tau_1 - \tau_2) + \cdots \qquad
\eea

We now consider the numerator of the spin-spin correlator in Eq.~(\ref{eqn:correlators}), given by the diagrams in Fig.~\ref{fig:2pnt1}.(a). 
\bea
\mathcal{Z} \mathcal{G}(\tau) &=& \Tr\left( \hS_{\alpha} \hS_{\beta} \right) + \Tr\left( \hS_{\alpha} \hS_{\sigma'} \hS_{\sigma'} \hS_{\beta} \right) \gamma_0^2 \int_0^{\tau} d\tau_1 \int_{\tau_1}^{\tau} d \tau_2 D_T(\tau_1 - \tau_2) \nn
&& + \ \Tr\left( \hS_{\alpha} \hS_{\beta} \hS_{\sigma'} \hS_{\sigma'} \right) \gamma_0^2 \int_{\tau}^{\beta} d\tau_1 \int_{\tau_1}^{\beta} d \tau_2 D_T(\tau_1 - \tau_2) \nn
&& + \ \Tr\left( \hS_{\alpha} \hS_{\sigma'} \hS_{\beta} \hS_{\sigma'} \right) \gamma_0^2 \int_0^{\tau} d\tau_1 \int_{\tau}^{\beta} d \tau_2 D_T(\tau_1 - \tau_2) + \cdots
\eea
Here, we take the external indices to either be $\alpha', \beta'$ to define $\mathcal{G}'(\tau)$, or $3$ to denote $\mathcal{G}_3(\tau)$. We notice that the three integrals contributing to the numerator are identical to the three we used to split up the denominator. Thus, to calculate the full correlation function, we only need to compute these three integrals and keep track of the difference in spin traces which appear in the numerator and denominator. This simplification is minor for the one-loop case, but it simplifies the two-loop calculation enormously. 

We now write the one-loop correlation function in terms of the spin traces given in Appendix \ref{traces}. Here, the traces on the right-hand side correspond to either the $\mathcal{S}'_i$ or $\mathcal{S}^z_i$ in the appendix depending on whether the left-hand side represents the correlator $\mathcal{G}'(\tau)$ or $\mathcal{G}^z(\tau)$ respectively.
\bea
\mathcal{G}(\tau) &=& \frac{S(S+1)}{3} \Bigg\{ 1 + \left[ \mathcal{S}_1 - \frac{2 S(S+1)}{3} \right] \gamma_0^2 \int_0^{\tau} d\tau_1 \int_{\tau_1}^{\tau} d \tau_2 D_T(\tau_1 - \tau_2) \nn
&& + \ \left[ \mathcal{S}_1 - \frac{2 S(S+1)}{3} \right] \gamma_0^2 \int_{\tau}^{\beta} d\tau_1 \int_{\tau_1}^{\beta} d \tau_2 D_T(\tau_1 - \tau_2) \nn
&& + \ \left[ \mathcal{S}_2 - \frac{2 S(S+1)}{3} \right] \gamma_0^2 \int_0^{\tau} d\tau_1 \int_{\tau}^{\beta} d \tau_2 D_T(\tau_1 - \tau_2) + \cdots \Bigg\}
\label{eqn:1loopg}
\eea

We now consider the evaluation of these integrals. For the purpose of renormalizing our theory, we can work in the $T = 0$ limit, where the bosonic propagator takes the form
\beq
D_0(\tau) = \int \frac{d^d k}{(2 \pi)^d} \frac{d \omega}{2 \pi} \frac{e^{- i \omega \tau}}{k^2 + \omega^2} = \frac{\tilde{S}_{d+1}}{|\tau|^{d - 1}}
\eeq
Finally, the integrations over imaginary time must be extended with care, since imaginary time is really compact: $\beta \sim 0$. Therefore, we need to extend the integration domain as
\beq
\int_0^{\beta} \longrightarrow \int_0^{\infty} + \int_{-\infty}^0
\eeq
so that the integration still forms a loop in imaginary time. So the three integrals appearing in Eq.~(\ref{eqn:1loopg}) respectively become
\beq
\int_0^{\tau} d\tau_1 \int_{\tau_1}^{\tau} d \tau_2 D_0(\tau_1 - \tau_2) = -\frac{\tilde{S}_{d+1} \tau^{\epsilon}}{\epsilon(1-\epsilon)}
\eeq
\beq
\left[ \int_{\tau}^{\infty} d\tau_1 \int_{\tau_1}^{\infty} d \tau_2 + \int_{\tau}^{\infty} d\tau_1 \int_{-\infty}^{0} d \tau_2 + \int_{-\infty}^{0} d\tau_1 \int_{\tau_1}^{0} d \tau_2 \right] D_0(\tau_1 - \tau_2) = -\frac{\tilde{S}_{d+1} \tau^{\epsilon}}{\epsilon(1-\epsilon)}
\eeq
\beq
\left[ \int_0^{\tau} d\tau_1 \int_{\tau}^{\infty} d \tau_2 + \int_0^{\tau} d\tau_1 \int_{-\infty}^{0} d \tau_2 \right] D_0(\tau_1 - \tau_2) = \frac{2 \tilde{S}_{d+1} \tau^{\epsilon}}{\epsilon(1-\epsilon)}
\eeq
where we have used the dimensional regularization ``identity'' $\int_0^{\infty} d\tau \tau^{\alpha} = 0$.

Collecting all of the above results, we find that the leading-order spin-spin correlation functions are given by
\beq
\mathcal{G}'(\tau) = \frac{S(S+1)}{3} \left[ 1 - \frac{\gamma_0^2 \tilde{S}_{d+1} \tau^{\epsilon}}{\epsilon(1-\epsilon)} + \cdots \right]
\eeq
\beq
\mathcal{G}^z(\tau) = \frac{S(S+1)}{3} \left[ 1 - \frac{2 \gamma_0^2 \tilde{S}_{d+1} \tau^{\epsilon}}{\epsilon(1-\epsilon)} + \cdots \right]
\eeq

The procedure at two-loop is done using the same procedure; we fill in the intermediate steps in Appendix \ref{sec:2loop}. Our final result is
\bea
\left\langle \hat{S}_{\alpha'}(\tau) \hat{S}_{\beta'}(0) \right\rangle &=& \delta_{\alpha' \beta'} \frac{S(S+1)}{3} \Bigg[ 1 - \frac{\gamma_0^2 \tilde{S}_{d+1} \tau^{\epsilon}}{\epsilon(1 - \epsilon)} \nn
&& + \ \left( \gamma_0^2 \tilde{S}_{d+1} \tau^{\epsilon} \right)^2 \bigg(\frac{1}{\epsilon^2} + \frac{5}{2 \epsilon} + \cdots \bigg) \Bigg]
\label{eqn:1loopg1}
\eea
\bea
\left\langle \hat{S}_{z}(\tau) \hat{S}_{z}(0) \right\rangle &=& \frac{S(S+1)}{3} \Bigg[ 1 - \frac{2 \gamma_0^2 \tilde{S}_{d+1} \tau^{\epsilon}}{\epsilon(1 - \epsilon)} \nn
&& + \ \left( \gamma_0^2 \tilde{S}_{d+1} \tau^{\epsilon} \right)^2 \bigg(\frac{3}{\epsilon^2} + \frac{6}{\epsilon} + \cdots \bigg) \Bigg]
\label{eqn:1loopg2}
\eea
where we only keep the divergent part of the $\gamma_0^4$ term.

\subsection{Vertex renormalization}

\begin{figure}
\includegraphics[width=15cm]{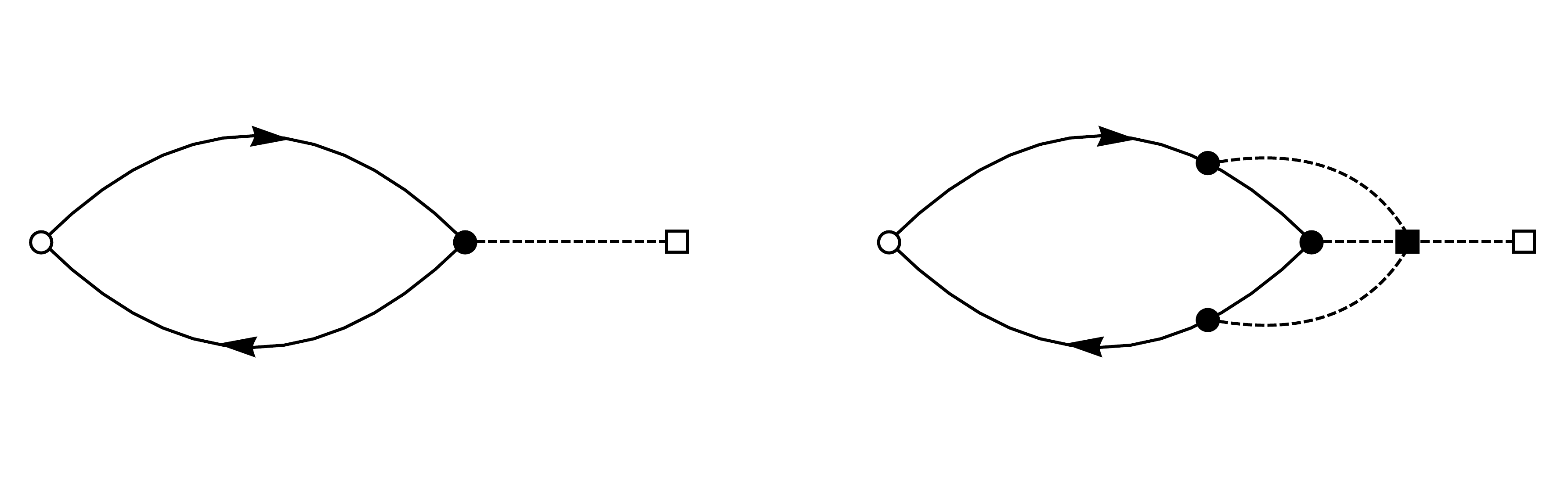}
\caption{The diagrams which renormalize the impurity interaction $\gamma$.}
\label{fig:vert}
\end{figure}

We now consider the renormalization of the vertex function $\mathcal{V}(x,\tau)$, defined in Eq.~(\ref{eqn:correlators}). In writing down all possible diagrams up to two-loop order, it becomes apparent that every diagram which does not depend on $u_0$ is identical to a diagram appearing in Fig.~\ref{fig:2pnt1}, but with the insertion of an external boson. Therefore, the only loop diagrams which contribute to renormalizing the bare interaction $\gamma_0$ are those which involve the bulk interaction; these are shown in Fig.~\ref{fig:vert}. This implies the \emph{exact} relation
\beq
Z_{\gamma} = 1 \quad \mrm{at} \quad g = 0
\eeq

We now evaluate the diagrams in Fig.~\ref{fig:vert} using the Feynman rules specified above. There is only one loop diagram which corrects the tree-level interaction, but there are three distinct ways to evaluate the spin traces. We find
\bea
\mathcal{V}(x,\tau) &=& \frac{S(S+1)}{3} \gamma_0 \int \frac{d^d k}{(2 \pi)^d} \frac{e^{ikx}}{k^2} \nn
&& - \  \frac{S(S+1)}{3} \frac{\gamma_0^3 u_0}{18} \int \frac{d^d k}{(2 \pi)^d} \frac{e^{ikx}}{k^2} \int \frac{d^d k_1}{(2 \pi)^d} \int \frac{d^d k_2}{(2 \pi)^d} \frac{2 \mathcal{S}'_1 + \mathcal{S}'_2}{k_1^2 k_2^2 (k + k_1 + k_2)^2}
\eea
where the spin traces $\mathcal{S}'_i$ are specified in Appendix \ref{traces}. Evaluating the divergent part of the integral, 
\bea
\mathcal{V}(x,\tau) &=& \frac{S(S+1)}{3} \int \frac{d^d k}{(2 \pi)^d} \frac{e^{ikx}}{k^2} \Bigg[ \gamma_0 \nn
&& - \ \gamma_0^3 u_0 \left( S(S+1) - \frac{1}{3} \right) \left(k^2\right)^{-\epsilon} \tilde{S}_{d+1}^2 \left( \frac{2 \pi^2}{15 \epsilon} + \cdots \right) \Bigg]
\label{eqn:vertren}
\eea

\section{Renormalization group summary}
\label{sec:rgsumm}

The RG equations can be obtained directly from the Eqns.~(\ref{eqn:1loopg1}), (\ref{eqn:1loopg2}), (\ref{eqn:vertren}), along with the definitions of the renormalization constants in Section \ref{sec:model}. After some algebra, we obtain
\bea
Z' &=& 1 - \frac{\gamma^2}{\epsilon} + \frac{1}{2 \epsilon} \gamma^4 \nn[.2cm]
Z_z &=& 1 - \frac{2 \gamma^2}{\epsilon} + \frac{1}{\epsilon^2}\gamma^4 \nn[.2cm]
Z_{\gamma} &=& 1 + \frac{2 \pi^2 \left[ S(S+1) - \frac{1}{3} \right] }{15 \epsilon} g \gamma^2
\eea
The beta function now follows from Eq.~(\ref{eqn:gammabeta}):
\bea
\beta_{\gamma} &=& - \frac{\epsilon}{2} \gamma + \frac{1}{2} \gamma^3 - \frac{1}{2} \gamma^5 + \frac{(N+2)}{144} g^2 \gamma + \ \frac{4 \pi^2}{15} \left[ S (S+1) - \frac{1}{3} \right]g \gamma^3
\eea
Tuning the bulk interactions to their fixed point, $g = g^{\ast}$, we find a fixed point for the impurity interactions which is also perturbative in $\epsilon$. To leading order,
\bea
\gamma^{\ast 2} &=& \epsilon + \bigg[ 1 - \frac{N+2}{2 (N+8)^2} - \ \frac{16 \pi^2}{5 (N+8)}\left( S (S+1) - \frac{1}{3} \right) \bigg] \epsilon^2
\eea
Since our model is symmetric under $\gamma \rightarrow -\gamma$, all physical quantities only depend on $\gamma^2$. The initial flow depends on the sign of the bare value of $\gamma_0$, after which the theory will flow to either $\gamma^{\ast}$ or $-\gamma^{\ast}$.

The anomalous dimensions of the spin operators follow from Eq.~(\ref{eqn:andim}):
\beq
\eta' = \gamma^2 - \gamma^4 
\eeq
\beq
\eta_z = 2 \gamma^2 
\eeq
where the $\mathcal{O}(\gamma^4)$ contribution to $\eta_z$ vanishes. Evaluating these at $\gamma = \gamma^{\ast}$:
\beq
\eta' = \epsilon - \left( \frac{N+2}{2 (N+8)^2} + \frac{16 \pi^2}{5 (N+8)}\left[ S (S+1) - \frac{1}{3} \right]  \right) \epsilon^2 
\eeq
\beq
\eta_z = 2 \epsilon + \bigg( 2 - \frac{N+2}{(N+8)^2} - \frac{32 \pi^2}{5 (N+8)}\left[ S (S+1) - \frac{1}{3} \right] \bigg) \epsilon^2
\eeq

As an aside, we mention the model with a Gaussian bulk, $g = 0$. This theory is infrared unstable to interactions, but the simple relation $Z_{\gamma} = 1$ allows us to derive an exact result for the anomalous dimension of the spin operators. Since the beta function for $\gamma$ only depends on $Z'$ in this theory, and $\beta_{\gamma} = 0$ at the interacting fixed point, Eqns.~(\ref{eqn:gammabeta})-(\ref{eqn:andim}) imply 
\beq
\eta' = \epsilon \quad \mrm{at} \quad g = 0
\eeq
to all orders in $\epsilon$. In contrast, $\eta_z$ will generically receive corrections at every order in $\epsilon$ at the Gaussian fixed point.

From Eqn.~(\ref{nudef}), we find the critical exponents
\beq
\nu' = 1 + \frac{\epsilon}{2} + \left( \frac{1}{4} - \frac{N+2}{4 (N+8)^2} - \frac{8 \pi^2}{5 (N+8)} \left[ S (S+1) - \frac{1}{3} \right] \right) \epsilon^2
\eeq
\beq
\nu_z = 1 + \epsilon + \left( 2 - \frac{N+2}{2 (N+8)^2} - \frac{16 \pi^2}{5 (N+8)} \left[ S (S+1) - \frac{1}{3} \right] \right) \epsilon^2
\eeq

We now compare these to numerical results. For $N=2$ and $S = 1/2$, we predict the critical exponents
\bea
\nu' &\approx& 1.08 \nn
\nu_z &\approx& 2.66
\eea

In Refs~\onlinecite{HCDS16,chen17}, both the microscopic model of Eqn.~(\ref{microH}) and the field theory model of Eqn.~(ref{spincoup}) were studied in numerical simulations. These authors claculated the above critical exponents to be
\bea 
\nu' &\approx& 1.13(2) \nn
\nu_z &\approx& 2.33(5)
\eea
The numerics show impressive agreement with the $\epsilon$ expansion.


\section{Compressibility}
\label{sec:compress}

In this section, we consider the finite-temperature response of the critical theory to an external probe coupled to the conserved $O(2)$ charge associated with particle number in the superfluid. Physically, this corresponds to the compressibility of the superfluid. We compute this by altering our Lagrangian,
\beq
\frac{1}{2} \int d^d x \left[\left(\partial_{\tau} \phi_1 \right)^2 + \left(\partial_{\tau} \phi_2 \right)^2\right] \longrightarrow \frac{1}{2}\int d^d x \left[ \left(\partial_{\tau} \phi_1 + i H \phi_2 \right)^2 + \left(\partial_{\tau} \phi_2 - i H \phi_1 \right)^2 \right] - H \hat{S}_z
\eeq
and then taking variational derivatives of the free energy
\beq
\chi = \frac{\delta^2 \left(T \log Z\right)}{\delta H^2}\Bigg|_{H = 0}
\label{eqn:funder}
\eeq
Here, we will continue working with our generalized theory, Eq.~(\ref{eq:model}), with $O(2) \times O(N - 2)$ symmetry, where the probe field $H$ couples to the $O(2)$ charge. The contribution of the bulk degrees of freedom to this quantity were computed in Ref.~\onlinecite{SS97}, so here we focus only on terms which depend on $\gamma$, and we denote this part of the compressibility by $\chi_{\mrm{imp}}$. Because this is a correlation function of a conserved current, its scaling dimension cannot renormalize, so at finite temperature it must take the form
\beq
\chi_{\mrm{imp}} = \frac{\mathcal{C}_1}{T}
\eeq
where $\mathcal{C}_1$ is a \emph{universal} number. We can also interpret $\mathcal{C}_1 = S_{\mrm{eff}}(S_{\mrm{eff}} + 1)/3$ as the ``effective spin'' in the presence of interactions with the bulk, since for $\gamma = 0$,
\beq
\chi_{\mrm{imp}}\bigg|_{\gamma = 0} = \frac{S(S+1)}{3T}
\eeq

In our calculations at $T=0$, we found that bulk interactions did not contribute to the impurity critical exponents until two-loop order. However, the structure of the $\epsilon$-expansion for the bulk theory is rather different at finite temperature. In the critical regime, physical quantities become an expansion in $\sqrt{\epsilon}$ (with possible extra factors of $\ln \epsilon$) \cite{SS97}. This dependence enters through the finite-temperature bosonic propagator, which is now given by
\beq
D_T(x,\tau) = T \sum_{i \omega_n} \int \frac{d^d k}{(2 \pi)^d} \frac{e^{i k x}e^{- i \omega_n \tau}}{\omega_n^2 + k^2 + m^2}
\eeq
with
\beq
m^2 = \left( \frac{N+2}{N+8} \right) \frac{2 \pi^2 T}{3} \epsilon
\eeq
We will see that this leads to a $\sqrt{\epsilon}$-expansion for $\chi_{\mrm{imp}}$ as well.

Performing the functional derivative in Eq.~(\ref{eqn:funder}), the compressibility is given by
\bea
\chi_{\mathrm{imp}} &=& \frac{1}{\beta} \int_0^{\beta} d\tau\int_0^{\beta} d\tau' \left\langle \hat{S}_z(\tau) \hat{S}_z(\tau') \right\rangle + \frac{1}{\beta} \int_0^{\beta} d \tau \int d^d x \ \left\langle \phi_{\alpha'}^2(\tau,x) \right\rangle \nn
&& - \ \frac{1}{\beta} \int_0^{\beta} d \tau \int_0^{\beta} d \tau' \int d^d x  \int d^d x' \left\langle \left[ \phi_2 \partial_{\tau} \phi_1 - \phi_1 \partial_{\tau} \phi_2 \right](\tau,x)\left[ \phi_2 \partial_{\tau'} \phi_1 - \phi_1 \partial_{\tau'} \phi_2 \right](\tau',x') \right\rangle \nn
&& - \frac{2 i}{\beta}\int_0^{\beta} d \tau \int_0^{\beta} d \tau' \int d^d x \left\langle \hat{S}_z(\tau) \left[ \phi_2 \partial_{\tau'} \phi_1 - \phi_1 \partial_{\tau'} \phi_2 \right](\tau',x) \right\rangle
\label{eqn:compres1}
\eea
These correlation functions can be computed using the same diagrammatic technique used in Section \ref{sec:renorm}, we simply do not take the zero temperature limit. A straight-forward computation leads to the expression
\bea
\chi_{\mathrm{imp}} &=& \frac{S(S+1)}{3 T} \left[ 1 + \frac{\gamma_0^2}{2 T} \int \frac{d^d k}{(2 \pi)^d} \frac{\mathrm{csch}^2\left( \beta \sqrt{k^2 + m^2}/2 \right)}{k^2 + m^2} \right]
\eea
Here we see why keeping the temperature-dependent mass in the bosonic propagator was crucial: for $m \rightarrow 0$ this expression in infrared singular, and an evaluation at finite $m$ gives (at leading order)
\beq
\chi_{\mathrm{imp}} = \frac{S(S+1)}{3 T} \left[ 1 + \gamma^2 \frac{\pi}{\beta m} \right]
\eeq
which lowers the order of the leading correction to
\beq
\chi_{\mathrm{imp}} = \frac{S(S+1)}{3 T} \left[ 1 + \left( \frac{3 (N+8)}{2(N+2)} \right)^{1/2} \sqrt{\epsilon} \right]
\eeq
As has been seen in previous work on the finite temperature $\epsilon$-expansion, the leading correction is not particularly small, so this may not give a good numerical estimate. For $S = 1/2$ and $N = 2$, we find
\beq
\mathcal{C}_1 \approx .734
\eeq

In Reference \onlinecite{chen17}, the constant $\mathcal{C}_1$ is computed numerically in a finite volume geometry, with a result close to the free value. Due to finite size effects, their result cannot be directly compared to ours.


\section{Conclusions}
\label{sec:summ}

Huang {\it et al.\/} \cite{HCDS16} recently found a novel impurity quantum critical point in their study of the Bose-Hubbard model on the square lattice. They held the bulk square lattice at the superfluid-insulator quantum critical point, and then varied the strength of the trapping potential at a single site. They found a quantum phase transition, with a diverging length scale, at a critical value of the trapping potential where the impurity site occupation number jumped by unity.

In an earlier study of quantum antiferromagnets with SU(2) spin rotation symmetry, Ref.~\onlinecite{SBV99} examined 
impurities in dimerized, two-dimensional antiferromagnets at the bulk critical point point between a spin-gap state and N\'eel order described by the O(3) Wilson-Fisher
conformal field theory. They found that impurities were universally characterized by a single spin quantum number, $S$, which specified a renormalization group fixed point with {\it no\/} relevant directions in the impurity field theory.

In this paper, we proposed that impurity criticality
of the Bose-Hubbard model \cite{HCDS16} is described by the $S=1/2$ impurity fixed point found in Ref.~\onlinecite{SBV99},
after the O(3) symmetry is reduced to O(2) in both the bulk and the impurity. We showed that with only O(2) symmetry, the impurity fixed point does allow for a {\it single\/} relevant perturbation in the impurity field theory: this relevant perturbation is associated with a longitudinal field acting on the $S=1/2$ spin on the impurity site. We note that a model of $S=1/2$ impurity has also been recently studied by Chen {\it et al.\/} \cite{chen17}.
With the presence of this relevant impurity perturbation, we can understand the need for a critical trapping potential in the numerical study of Huang {\it et al.\/} \cite{HCDS16}.

We computed critical exponents and universal amplitudes associated with the O(2)-symmetric impurity fixed point in an expansion in $\epsilon=3-d$, where $d$ is the bulk spatial dimensionality. Associated with two different relevant perturbations, we estimated from a computation to order
$\epsilon^2$ that the impurity length scale diverged with the exponents $\nu_z \approx 2.66$ and $\nu' \approx 1.08$; this compares well with the numerical results \cite{HCDS16,chen17} $\nu_z \approx 2.33$ and $\nu' \approx 1.13$.
Additional tests of the $\epsilon$-expansion results will be possible in further numerical studies.

Finally, we note that this novel impurity quantum criticality should be accessible in cold atom experiments, and we hope it will be studied in the near future.

\section*{Acknowledgements}

We thank K. Chen and B. Svistunov for helpful discussions, and for sharing their preliminary results with us. This research was supported by the NSF under Grants DMR-1360789 and DMR-1664842, and the MURI grant W911NF-14-1-0003 from ARO. Research at Perimeter Institute is supported by the Government of Canada through Industry Canada and by the Province of Ontario through the Ministry of Research and Innovation. SS also acknowledges support from Cenovus Energy at Perimeter Institute. 

\appendix
\section{Spin traces}
\label{traces}

Here we tabulate spin traces. We give expressions in terms of the index $\mathcal{C}_S$ of the spin-$S$ representation of $SU(2)$,
\beq
\mathcal{C}_S = \frac{1}{3} (2S+1) S (S+1)
\eeq
This is defined as the constant appearing in the bilinear trace
\beq
\Tr \left( \hat{S}_{\alpha} \hat{S}_{\beta} \right) = \mathcal{C}_S \delta_{\alpha \beta},
\eeq

Below we give the relevant traces, where we distinguish $\sigma' = 1,2$ from the $z = 3$ direction. These traces give zero if one replaces one of the two $\sigma'$ indices with $z$.

At one-loop, we need the following traces:
\bea
\Tr \left( \hat{S}_{\alpha'} \hat{S}_{\sigma'} \hat{S}_{\sigma'} \hat{S}_{\beta'} \right) &=& \left[ \frac{4}{5} S (S+1) - \frac{1}{10} \right] \mathcal{C}_S \delta_{\alpha' \beta'} = \mathcal{S}'_1 \mathcal{C}_S \delta_{\alpha' \beta'} \nn
\Tr \left( \hat{S}_{z} \hat{S}_{\sigma'} \hat{S}_{\sigma'} \hat{S}_{z} \right) &=& \left[ \frac{2}{5} S (S+1) + \frac{1}{5} \right] \mathcal{C}_S = \mathcal{S}^z_1 \mathcal{C}_S \nn
\Tr \left( \hat{S}_{\alpha'} \hat{S}_{\sigma'} \hat{S}_{\beta'} \hat{S}_{\sigma'}  \right) &=& \left[ \frac{4}{5} S (S+1) - \frac{3}{5} \right] \mathcal{C}_S \delta_{\alpha' \beta'} = \mathcal{S}'_2 \mathcal{C}_S \delta_{\alpha' \beta'} \nn
\Tr\left( \hat{S}_{z} \hat{S}_{\sigma'} \hat{S}_{z} \hat{S}_{\sigma'}  \right) &=& \left[ \frac{2}{5} S (S+1) - \frac{4}{5} \right] \mathcal{C}_S = \mathcal{S}^z_2 \mathcal{C}_S
\eea

At two-loop:
\bea
\Tr \left( \hat{S}_{\alpha'} \hat{S}_{\sigma'} \hat{S}_{\sigma'} \hat{S}_{\eta'} \hat{S}_{\eta'} \hat{S}_{\beta'} \right) &=& \frac{\mathcal{C}_S}{70} \left[ 48 \ S^2(S+1)^2- 6 S(S+1) - 5 \right] \delta_{\alpha' \beta'} = \mathcal{S}'_3 \mathcal{C}_S \delta_{\alpha' \beta'} \nn
\Tr \left( \hat{S}_{z} \hat{S}_{\sigma'} \hat{S}_{\sigma'} \hat{S}_{\eta'} \hat{S}_{\eta'} \hat{S}_{z} \right) &=& \frac{\mathcal{C}_S}{35} \left[ 8 \ S^2(S+1)^2- S(S+1) + 5 \right] = \mathcal{S}^z_3 \mathcal{C}_S \nn
\Tr \left( \hat{S}_{\alpha'} \hat{S}_{\sigma'} \hat{S}_{\sigma'} \hat{S}_{\eta'} \hat{S}_{\beta'} \hat{S}_{\eta'} \right) &=& \frac{\mathcal{C}_S}{70} \left[ 48 \ S^2(S+1)^2- 48 S(S+1) + 9 \right] \delta_{\alpha' \beta'} = \mathcal{S}'_4 \mathcal{C}_S \delta_{\alpha' \beta'} \nn
\Tr \left( \hat{S}_{z} \hat{S}_{\sigma'} \hat{S}_{\sigma'} \hat{S}_{\eta'} \hat{S}_{z} \hat{S}_{\eta'} \right) &=& \frac{\mathcal{C}_S}{35} \left[ 8 \ S^2(S+1)^2 - 15 S(S+1) - 2 \right] = \mathcal{S}^z_4 \mathcal{C}_S \nn
\Tr \left( \hat{S}_{\alpha'} \hat{S}_{\sigma'} \hat{S}_{\sigma'} \hat{S}_{\beta'} \hat{S}_{\eta'} \hat{S}_{\eta'} \right) &=& \frac{\mathcal{C}_S}{35} \left[ 24 \ S^2(S+1)^2- 17 S(S+1) + 8 \right] \delta_{\alpha' \beta'} = \mathcal{S}'_5 \mathcal{C}_S \delta_{\alpha' \beta'} \nn
\Tr \left( \hat{S}_{z} \hat{S}_{\sigma'} \hat{S}_{\sigma'} \hat{S}_{z} \hat{S}_{\eta'} \hat{S}_{\eta'} \right) &=& \frac{\mathcal{C}_S}{35} \left[ 8 \ S^2(S+1)^2- S(S+1) + 5 \right] = \mathcal{S}^z_5 \mathcal{C}_S \nn
\Tr \left( \hat{S}_{\alpha'} \hat{S}_{\sigma'} \hat{S}_{\eta'} \hat{S}_{\eta'} \hat{S}_{\sigma'} \hat{S}_{\beta'} \right) &=& \frac{\mathcal{C}_S}{35} \left[ 24 \ S^2(S+1)^2 - 17 S(S+1) + 8 \right] \delta_{\alpha' \beta'} = \mathcal{S}'_6 \mathcal{C}_S \delta_{\alpha' \beta'} \nn
\Tr \left( \hat{S}_{z} \hat{S}_{\sigma'} \hat{S}_{\eta'} \hat{S}_{\eta'} \hat{S}_{\sigma'} \hat{S}_{z} \right) &=& \frac{\mathcal{C}_S}{35} \left[ 8 \ S^2(S+1)^2 + 27 S(S+1) - 16 \right] = \mathcal{S}^z_6 \mathcal{C}_S \nn
\Tr \left( \hat{S}_{\alpha'} \hat{S}_{\sigma'} \hat{S}_{\eta'} \hat{S}_{\beta'} \hat{S}_{\eta'} \hat{S}_{\sigma'} \right) &=& \frac{\mathcal{C}_S}{35} \left[ 24 \ S^2(S+1)^2 - 38 S(S+1) + 15 \right] \delta_{\alpha' \beta'} = \mathcal{S}'_7 \mathcal{C}_S \delta_{\alpha' \beta'} \nn
\Tr \left( \hat{S}_{z} \hat{S}_{\sigma'} \hat{S}_{\eta'} \hat{S}_{z} \hat{S}_{\eta'} \hat{S}_{\sigma'} \right) &=& \frac{\mathcal{C}_S}{35} \left[ 8 \ S^2(S+1)^2 - 29 S(S+1) + 26 \right] = \mathcal{S}^z_7 \mathcal{C}_S \nn
\Tr \left( \hat{S}_{\alpha'} \hat{S}_{\sigma'} \hat{S}_{\eta'} \hat{S}_{\sigma'} \hat{S}_{\eta'} \hat{S}_{\beta'} \right) &=& \frac{\mathcal{C}_S}{70} \left[ 48 \ S^2(S+1)^2- 48 S(S+1) + 9 \right] \delta_{\alpha' \beta'} = \mathcal{S}'_8 \mathcal{C}_S \delta_{\alpha' \beta'} \nn
\Tr \left( \hat{S}_{z} \hat{S}_{\sigma'} \hat{S}_{\eta'} \hat{S}_{\sigma'} \hat{S}_{\eta'} \hat{S}_{z} \right) &=& \frac{\mathcal{C}_S}{35} \left[ 8 \ S^2(S+1)^2 + 6 S(S+1) - 9 \right] = \mathcal{S}^z_8 \mathcal{C}_S \nn
\Tr \left( \hat{S}_{\alpha'} \hat{S}_{\sigma'} \hat{S}_{\eta'} \hat{S}_{\sigma'} \hat{S}_{\beta'} \hat{S}_{\eta'} \right) &=& \frac{\mathcal{C}_S}{35} \left[ 24 \ S^2(S+1)^2- 38 S(S+1) + 15 \right] \delta_{\alpha' \beta'} = \mathcal{S}'_9 \mathcal{C}_S \delta_{\alpha' \beta'} \nn
\Tr \left( \hat{S}_{z} \hat{S}_{\sigma'} \hat{S}_{\eta'} \hat{S}_{\sigma'} \hat{S}_{z} \hat{S}_{\eta'} \right) &=& \frac{\mathcal{C}_S}{35} \left[ 8 \ S^2(S+1)^2 - 22 S(S+1) + 12 \right] = \mathcal{S}^z_9 \mathcal{C}_S \nn
\Tr \left( \hat{S}_{\alpha'} \hat{S}_{\sigma'} \hat{S}_{\eta'} \hat{S}_{\beta'} \hat{S}_{\sigma'} \hat{S}_{\eta'} \right) &=& \frac{\mathcal{C}_S}{35} \left[ 24 \ S^2(S+1)^2 - 45 S(S+1) + 29 \right] \delta_{\alpha' \beta'} = \mathcal{S}'_{10} \mathcal{C}_S \delta_{\alpha' \beta'} \nn
\Tr \left( \hat{S}_{z} \hat{S}_{\sigma'} \hat{S}_{\eta'} \hat{S}_{z} \hat{S}_{\sigma'} \hat{S}_{\eta'} \right) &=& \frac{\mathcal{C}_S}{35} \left[ 8 \ S^2(S+1)^2 - 50 S(S+1) + 33 \right] = \mathcal{S}^z_{10} \mathcal{C}_S
\eea

\section{Details of the two-loop calculation}
\label{sec:2loop}

In this appendix we detail some of the intermediate steps in the calculation of the two-loop contribution to the spin-spin correlation function quoted in Eqns.~(\ref{eqn:1loopg1})-(\ref{eqn:1loopg2}). The procedure proceeds in a very similar fashion to how the one-loop calculation is described in Section \ref{sec:2pnt1}.

\begin{figure}
\includegraphics[width=15cm]{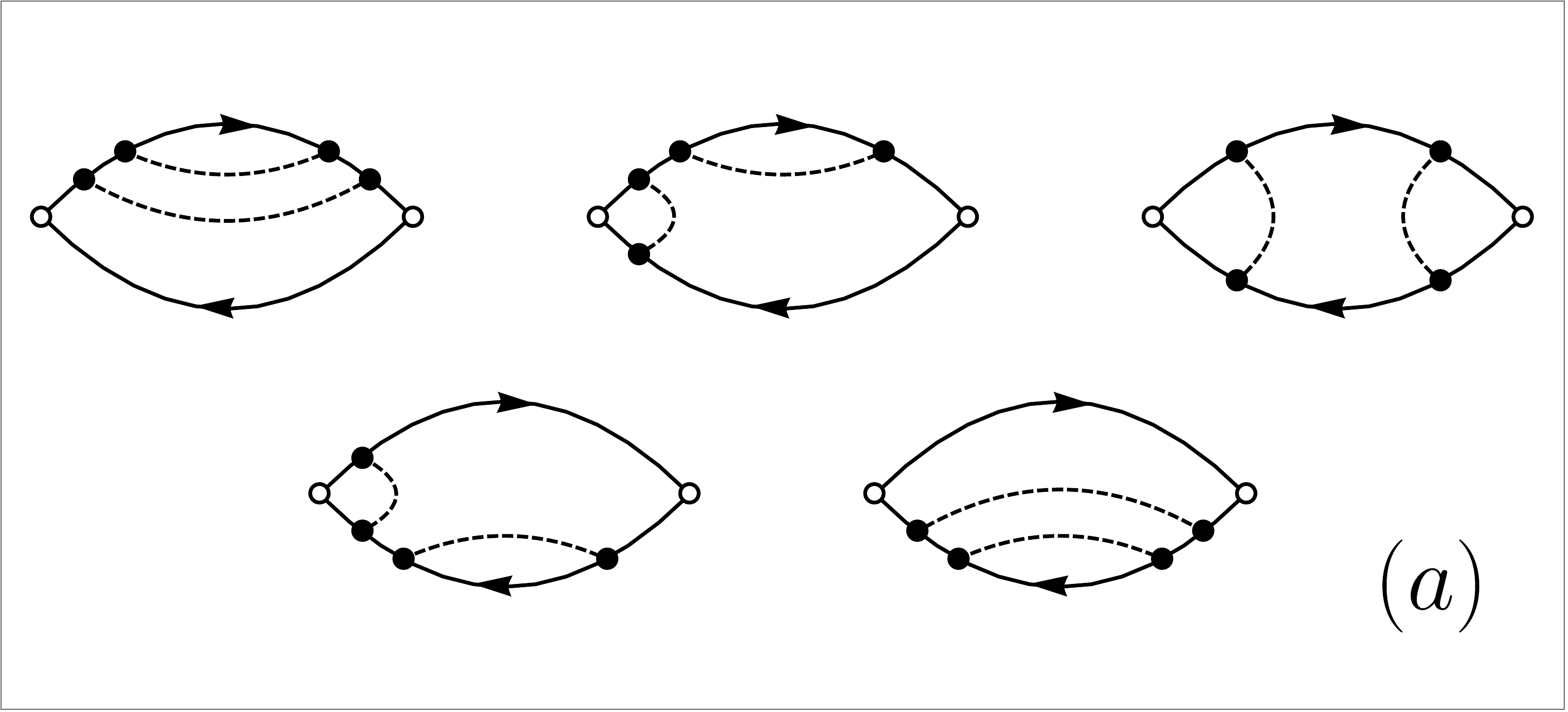}
\includegraphics[width=15cm]{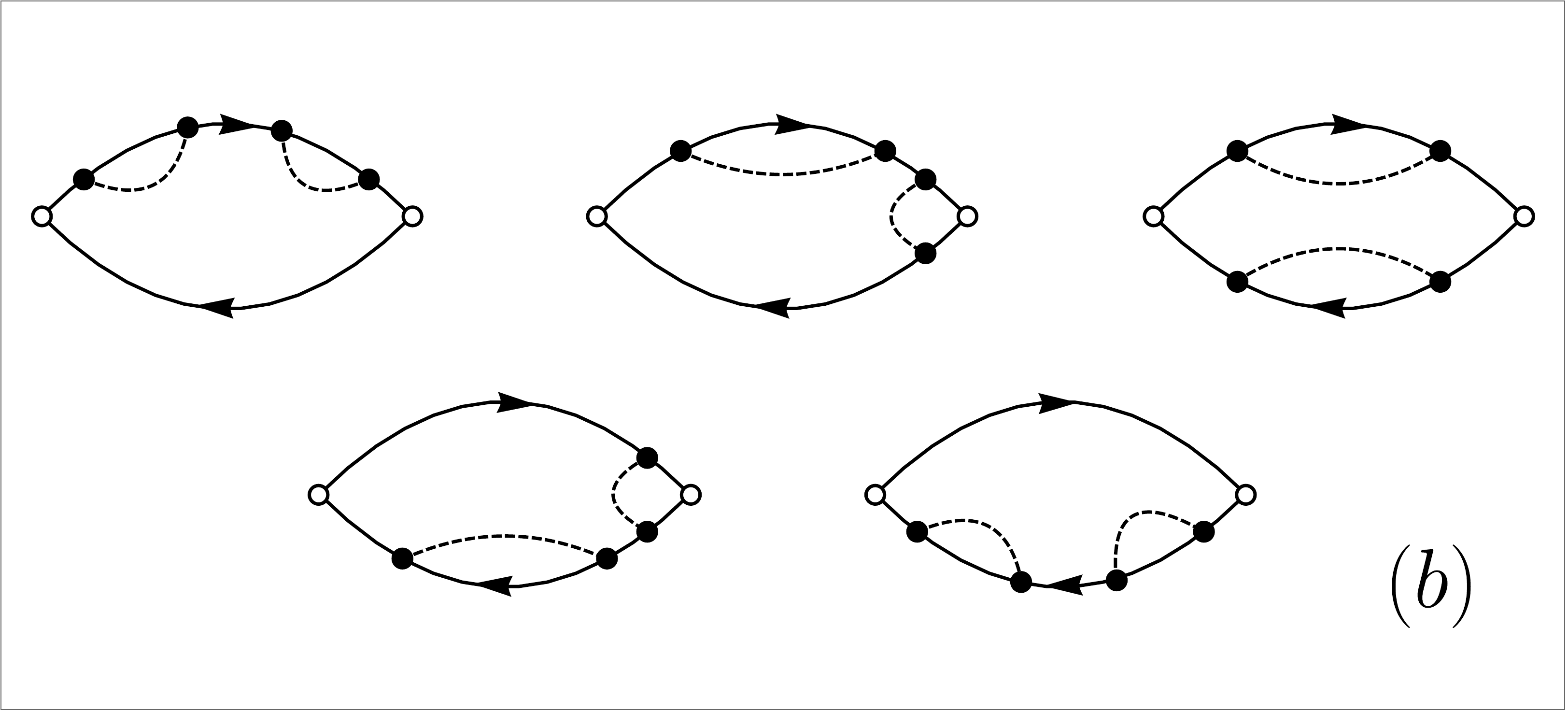}
\includegraphics[width=15cm]{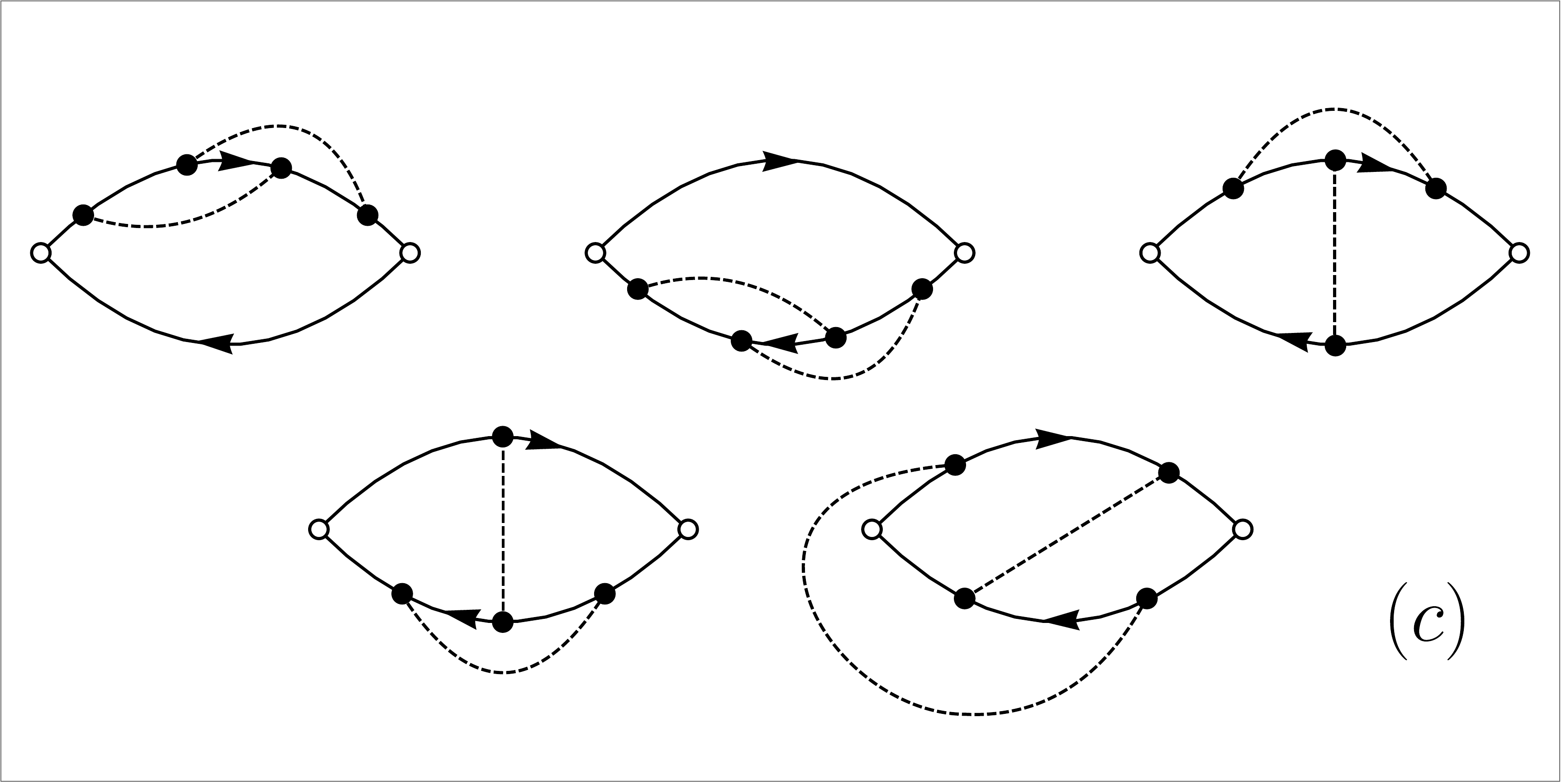}
\caption{The diagrams contributing to the numerator of the two-point function at two-loop.}
\label{fig:2loop}
\end{figure}

\begin{figure}
\includegraphics[width=14cm]{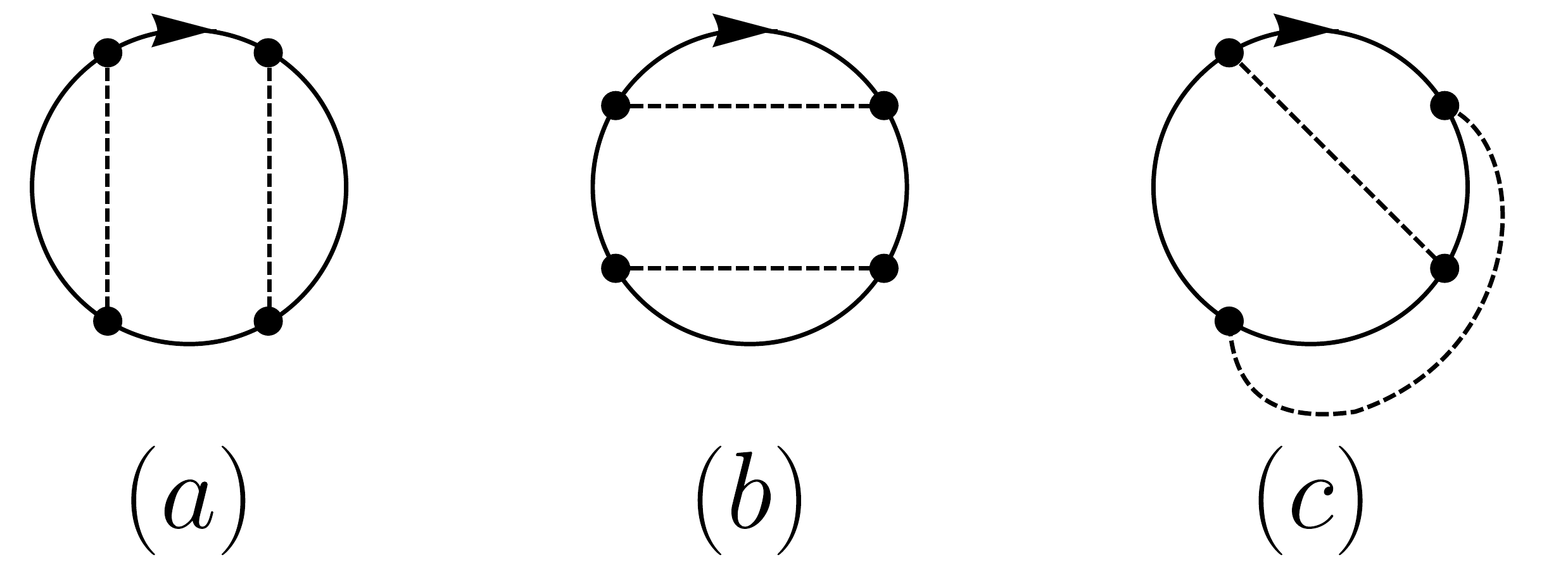}
\caption{The diagrams contributing to the denominator of the two-point function at two-loop.}
\label{fig:z2loop}
\end{figure}

The relevant diagrams are pictured in Figures \ref{fig:2loop} and \ref{fig:z2loop}. Here, we have grouped the diagrams into three groups (a), (b), and (c). This is because, like the one-loop calculation in the main text, the three diagrams contributing to the denominator can be rewritten so that they are the sum of the diagrams in the numerator. Then we only need to compute the 15 diagrams which follow form the integrals pictured in Figure \ref{fig:2loop}, while keeping track of the difference in spin traces between the numerator and denominator. In the $O(3)$ symmetric case considered in Reference \onlinecite{VBS00}, this resulted in large cancellations and only 7 diagrams need to be computed. In contrast, there are no cancellations here, and all 15 diagrams need to calculated.

We label the loop integrals which follow from Figure \ref{fig:2loop} as $\mathcal{I}_i$ for $i = 1, 2, ..., 15$, where we label the integrals from left-to-right and top-to-bottom according to the figure. In terms of these integrals, the two-loop contribution to $\mathcal{G}$ is
\bea
\mathcal{G}^{(\mrm{two-loop})} &=& \frac{S(S+1)}{3} \gamma_0^4 \Bigg\{ \left[ \mathcal{S}_6 - \frac{2 S(S+1)}{3}\mathcal{S}'_1 \right] \mathcal{I}_1 + \left[ \mathcal{S}_4 - \frac{2 S(S+1)}{3}\mathcal{S}'_1 \right] \mathcal{I}_2 \nn
&& + \ \left[ \mathcal{S}_7 - \frac{2 S(S+1)}{3}\mathcal{S}'_1 \right] \mathcal{I}_3 + \left[ \mathcal{S}_4 - \frac{2 S(S+1)}{3}\mathcal{S}'_1 \right] \mathcal{I}_4 \nn
&& + \ \left[ \mathcal{S}_6 - \frac{2 S(S+1)}{3}\mathcal{S}'_1 \right] \mathcal{I}_5  + \left[ \mathcal{S}_3 - \frac{2 S(S+1)}{3}\mathcal{S}'_1 \right] \mathcal{I}_6 \nn
&& + \ \left[ \mathcal{S}_4 - \frac{2 S(S+1)}{3}\mathcal{S}'_1 \right] \mathcal{I}_7 + \left[ \mathcal{S}_5 - \frac{2 S(S+1)}{3}\mathcal{S}'_1 \right] \mathcal{I}_8 \nn
&& + \ \left[ \mathcal{S}_4 - \frac{2 S(S+1)}{3}\mathcal{S}'_1 \right] \mathcal{I}_9 + \left[ \mathcal{S}_3 - \frac{2 S(S+1)}{3}\mathcal{S}'_1 \right] \mathcal{I}_{10} \nn
&& + \left[ \mathcal{S}_8 - \frac{2 S(S+1)}{3}\mathcal{S}'_2 \right] \mathcal{I}_{11} + \left[ \mathcal{S}_8 - \frac{2 S(S+1)}{3}\mathcal{S}'_2 \right] \mathcal{I}_{12} \nn
&& + \ \left[ \mathcal{S}_9 - \frac{2 S(S+1)}{3}\mathcal{S}'_2 \right] \mathcal{I}_{13} + \left[ \mathcal{S}_9 - \frac{2 S(S+1)}{3}\mathcal{S}'_2 \right] \mathcal{I}_{14} \nn
&& + \ \left[ \mathcal{S}_{10} - \frac{2 S(S+1)}{3}\mathcal{S}'_2 \right] \mathcal{I}_{15} \Bigg\}
\label{eqn:g2loop}
\eea
Within each bracket, the first spin sum is either $\mathcal{S}'$ or $\mathcal{S}^z$ depending on whether one wants the two point correlator $\mathcal{G}'$ or $\mathcal{G}^z$. We note that the denominator $Z$ also contains an order $\gamma_0^4$ term from expanding the one-loop contribution to second order, but this contribution vanishes in dimensional regularization.

We now evaluate the 15 integrals above. Below, we will give the $T > 0$ integrals for each integral which follow from the diagrams in Fig.~\ref{fig:2loop}, and then state the evaluation of the divergent piece of the $T = 0$ limit. We take this limit according to the prescription described below Eq.~\ref{eqn:1loopg} in the main text. 

\bea
\mathcal{I}_1 &=& \int_0^{\tau} d \tau_1 \int_{\tau_1}^{\tau} d \tau_2 \int_{\tau_1}^{\tau_2} d \tau_3 \int_{\tau_3}^{\tau_2} d \tau_4 D(\tau_1 - \tau_2) D(\tau_3 - \tau_4) \nn
&\overset{\beta \rightarrow \infty}{\Longrightarrow}& \ \left( \tilde{S}_{d+1} \tau^{\epsilon} \right)^2 \left( \frac{1}{2 \epsilon^2} + \frac{3}{2 \epsilon} + \cdots \right)
\eea
\bea
\mathcal{I}_2 &=& \int_0^{\tau} d \tau_1 \int_{\tau}^{\beta} d \tau_2 \int_{\tau_1}^{\tau} d \tau_3 \int_{\tau_3}^{\tau} d \tau_4 D(\tau_1 - \tau_2) D(\tau_3 - \tau_4) \nn
&\overset{\beta \rightarrow \infty}{\Longrightarrow}& \ \left( \tilde{S}_{d+1} \tau^{\epsilon} \right)^2 \left( -\frac{3}{2 \epsilon^2} - \frac{3}{\epsilon} + \cdots \right)
\eea
\bea
\mathcal{I}_3 &=& \int_0^{\tau} d \tau_1 \int_{\tau}^{\beta} d \tau_2 \int_{\tau_1}^{\tau} d \tau_3 \int_{\tau}^{\tau_2} d \tau_4 D(\tau_1 - \tau_2) D(\tau_3 - \tau_4) \nn
&\overset{\beta \rightarrow \infty}{\Longrightarrow}& \ \left( \tilde{S}_{d+1} \tau^{\epsilon} \right)^2 \left( \frac{2}{\epsilon^2} + \frac{3}{\epsilon} + \cdots \right)
\eea
\bea
\mathcal{I}_4 &=& \int_0^{\tau} d \tau_1 \int_{\tau}^{\beta} d \tau_2 \int_{\tau}^{\tau_2} d \tau_3 \int_{\tau}^{\tau_3} d \tau_4 D(\tau_1 - \tau_2) D(\tau_3 - \tau_4) \nn
&\overset{\beta \rightarrow \infty}{\Longrightarrow}& \ \left( \tilde{S}_{d+1} \tau^{\epsilon} \right)^2 \left( -\frac{3}{2 \epsilon^2} - \frac{3}{\epsilon} + \cdots \right)
\eea
\bea
\mathcal{I}_5 &=& \int_{\tau}^{\beta} d \tau_1 \int_{\tau}^{\tau_1} d \tau_2 \int_{\tau_2}^{\tau_1} d \tau_3 \int_{\tau_2}^{\tau_3} d \tau_4 D(\tau_1 - \tau_2) D(\tau_3 - \tau_4) \nn
&\overset{\beta \rightarrow \infty}{\Longrightarrow}& \ \left( \tilde{S}_{d+1} \tau^{\epsilon} \right)^2 \left( \frac{1}{2 \epsilon^2} + \frac{3}{2\epsilon} + \cdots \right)
\eea

\bea
\mathcal{I}_6 &=& \int_0^{\tau} d \tau_1 \int_{\tau_1}^{\tau} d \tau_2 \int_{\tau_2}^{\tau} d \tau_3 \int_{\tau_3}^{\tau} d \tau_4 D(\tau_1 - \tau_2) D(\tau_3 - \tau_4) \nn
&\overset{\beta \rightarrow \infty}{\Longrightarrow}& \ \left( \tilde{S}_{d+1} \tau^{\epsilon} \right)^2 \left( \frac{1}{\epsilon^2} + \frac{2}{\epsilon} + \cdots \right)
\eea
\bea
\mathcal{I}_7 &=& \int_0^{\tau} d \tau_1 \int_{\tau_1}^{\tau} d \tau_2 \int_{\tau_2}^{\tau} d \tau_3 \int_{\tau}^{\beta} d \tau_4 D(\tau_1 - \tau_2) D(\tau_3 - \tau_4) \nn
&\overset{\beta \rightarrow \infty}{\Longrightarrow}& \ \left( \tilde{S}_{d+1} \tau^{\epsilon} \right)^2 \left( - \frac{3}{2 \epsilon^2} - \frac{3}{\epsilon} + \cdots \right)
\eea
\bea
\mathcal{I}_8 &=& \int_0^{\tau} d \tau_1 \int_{\tau_1}^{\tau} d \tau_2 \int_{\tau}^{\beta} d \tau_3 \int_{\tau_3}^{\beta} d \tau_4 D(\tau_1 - \tau_2) D(\tau_3 - \tau_4) \nn
&\overset{\beta \rightarrow \infty}{\Longrightarrow}& \ \left( \tilde{S}_{d+1} \tau^{\epsilon} \right)^2 \left( \frac{1}{\epsilon^2} + \frac{2}{\epsilon} + \cdots \right)
\eea
\bea
\mathcal{I}_9 &=& \int_0^{\tau} d \tau_1 \int_{\tau}^{\beta} d \tau_2 \int_{\tau_2}^{\beta} d \tau_3 \int_{\tau_3}^{\beta} d \tau_4 D(\tau_1 - \tau_2) D(\tau_3 - \tau_4) \nn
&\overset{\beta \rightarrow \infty}{\Longrightarrow}& \ \left( \tilde{S}_{d+1} \tau^{\epsilon} \right)^2 \left( - \frac{3}{2 \epsilon^2} - \frac{3}{\epsilon} + \cdots \right)
\eea
\bea
\mathcal{I}_{10} &=& \int_{\tau}^{\beta} d \tau_1 \int_{\tau_1}^{\beta} d \tau_2 \int_{\tau_2}^{\beta} d \tau_3 \int_{\tau_3}^{\beta} d \tau_4 D(\tau_1 - \tau_2) D(\tau_3 - \tau_4) \nn
&\overset{\beta \rightarrow \infty}{\Longrightarrow}& \ \left( \tilde{S}_{d+1} \tau^{\epsilon} \right)^2 \left( \frac{1}{\epsilon^2} + \frac{2}{\epsilon} + \cdots \right)
\eea

\bea
\mathcal{I}_{11} &=& \int_0^{\tau} d \tau_1 \int_{\tau_1}^{\tau} d \tau_2 \int_{\tau_1}^{\tau_2} d \tau_3 \int_{\tau_2}^{\tau} d \tau_4 D(\tau_1 - \tau_2) D(\tau_3 - \tau_4) \nn
&\overset{\beta \rightarrow \infty}{\Longrightarrow}& \left( \tilde{S}_{d+1} \tau^{\epsilon} \right)^2 \left( -\frac{1}{\epsilon^2} - \frac{5}{2\epsilon} \right) 
\eea
\bea
\mathcal{I}_{12} &=& \int_{\tau}^{\beta} d \tau_1 \int_{\tau}^{\tau_1} d \tau_2 \int_{\tau_2}^{\tau_1} d \tau_3 \int_{\tau}^{\tau_2} d \tau_4 D(\tau_1 - \tau_2) D(\tau_3 - \tau_4) \nn
&\overset{\beta \rightarrow \infty}{\Longrightarrow}& \left( \tilde{S}_{d+1} \tau^{\epsilon} \right)^2 \left( -\frac{1}{\epsilon^2} - \frac{5}{2\epsilon} \right) 
\eea
\bea
\mathcal{I}_{13} &=& \int_{0}^{\tau} d \tau_1 \int_{\tau}^{\beta} d \tau_2 \int_{\tau}^{\tau_2} d \tau_3 \int_{\tau_2}^{\beta} d \tau_4 D(\tau_1 - \tau_2) D(\tau_3 - \tau_4) \nn
&\overset{\beta \rightarrow \infty}{\Longrightarrow}& \left( \tilde{S}_{d+1} \tau^{\epsilon} \right)^2 \left(\frac{1}{\epsilon^2} + \frac{2}{\epsilon} + \cdots\right)
\eea
\bea
\mathcal{I}_{14} &=& \int_{0}^{\tau} d \tau_1 \int_{\tau}^{\beta} d \tau_2 \int_{0}^{\tau_1} d \tau_3 \int_{\tau_1}^{\tau} d \tau_4 D(\tau_1 - \tau_2) D(\tau_3 - \tau_4) \nn
&\overset{\beta \rightarrow \infty}{\Longrightarrow}& \left( \tilde{S}_{d+1} \tau^{\epsilon} \right)^2 \left(\frac{1}{\epsilon^2} + \frac{2}{\epsilon} + \cdots\right)
\eea
\bea
\mathcal{I}_{15} &=& \int_{0}^{\tau} d \tau_1 \int_{\tau}^{\beta} d \tau_2 \int_{0}^{\tau_1} d \tau_3 \int_{\tau}^{\tau_2} d \tau_4 D(\tau_1 - \tau_2) D(\tau_3 - \tau_4) \nn
&\overset{\beta \rightarrow \infty}{\Longrightarrow}& \left( \tilde{S}_{d+1} \tau^{\epsilon} \right)^2 \left(\frac{1}{\epsilon} + \cdots\right)
\eea

Plugging these values into Eq.~(\ref{eqn:g2loop}) and simplifying gives the full two-loop expression used in Eq.~(\ref{eqn:1loopg2}) in the main text.

\bibliography{impurity}
\end{document}